\documentclass[preprint,5p,twocolumn,sort&compress,times]{elsarticle}

\usepackage{amssymb}

\usepackage{lineno}

\usepackage{url}
\usepackage{amsmath}

\journal{Journal of High Energy Astrophysics}

\def\CRBeam{{\it CRBeam}}
\def\Fermitools{{\it Fermitools}}
\def\FermiPy{{\it FermiPy}}
\def\iminuit{{\it iminuit}}

\makeatletter

\let\jnl@style=\rm
\def\ref@jnl#1{{\jnl@style#1}}

\def\aj{\ref@jnl{AJ}}                   
\def\actaa{\ref@jnl{Acta Astron.}}      
\def\araa{\ref@jnl{ARA\&A}}             
\def\apj{\ref@jnl{ApJ}}                 
\def\apjl{\ref@jnl{ApJ}}                
\def\apjs{\ref@jnl{ApJS}}               
\def\ao{\ref@jnl{Appl.~Opt.}}           
\def\apss{\ref@jnl{Ap\&SS}}             
\def\aap{\ref@jnl{A\&A}}                
\def\aapr{\ref@jnl{A\&A~Rev.}}          
\def\aaps{\ref@jnl{A\&AS}}              
\def\azh{\ref@jnl{AZh}}                 
\def\baas{\ref@jnl{BAAS}}               
\def\bac{\ref@jnl{Bull. astr. Inst. Czechosl.}}
\def\caa{\ref@jnl{Chinese Astron. Astrophys.}}
\def\cjaa{\ref@jnl{Chinese J. Astron. Astrophys.}}
\def\icarus{\ref@jnl{Icarus}}           
\def\jcap{\ref@jnl{J. Cosmology Astropart. Phys.}}
\def\jrasc{\ref@jnl{JRASC}}             
\def\memras{\ref@jnl{MmRAS}}            
\def\mnras{\ref@jnl{MNRAS}}             
\def\na{\ref@jnl{New A}}                
\def\nar{\ref@jnl{New A Rev.}}          
\def\pra{\ref@jnl{Phys.~Rev.~A}}        
\def\prb{\ref@jnl{Phys.~Rev.~B}}        
\def\prc{\ref@jnl{Phys.~Rev.~C}}        
\def\prd{\ref@jnl{Phys.~Rev.~D}}        
\def\pre{\ref@jnl{Phys.~Rev.~E}}        
\def\prl{\ref@jnl{Phys.~Rev.~Lett.}}    
\def\pasa{\ref@jnl{PASA}}               
\def\pasp{\ref@jnl{PASP}}               
\def\pasj{\ref@jnl{PASJ}}               
\def\rmxaa{\ref@jnl{Rev. Mexicana Astron. Astrofis.}}%
\def\qjras{\ref@jnl{QJRAS}}             
\def\skytel{\ref@jnl{S\&T}}             
\def\solphys{\ref@jnl{Sol.~Phys.}}      
\def\sovast{\ref@jnl{Soviet~Ast.}}      
\def\ssr{\ref@jnl{Space~Sci.~Rev.}}     
\def\zap{\ref@jnl{ZAp}}                 
\def\nat{\ref@jnl{Nature}}              
\def\iaucirc{\ref@jnl{IAU~Circ.}}       
\def\aplett{\ref@jnl{Astrophys.~Lett.}} 
\def\apspr{\ref@jnl{Astrophys.~Space~Phys.~Res.}}
\def\bain{\ref@jnl{Bull.~Astron.~Inst.~Netherlands}} 
\def\fcp{\ref@jnl{Fund.~Cosmic~Phys.}}  
\def\gca{\ref@jnl{Geochim.~Cosmochim.~Acta}}   
\def\grl{\ref@jnl{Geophys.~Res.~Lett.}} 
\def\jcp{\ref@jnl{J.~Chem.~Phys.}}      
\def\jgr{\ref@jnl{J.~Geophys.~Res.}}    
\def\jqsrt{\ref@jnl{J.~Quant.~Spec.~Radiat.~Transf.}}
\def\memsai{\ref@jnl{Mem.~Soc.~Astron.~Italiana}}
\def\nphysa{\ref@jnl{Nucl.~Phys.~A}}   
\def\physrep{\ref@jnl{Phys.~Rep.}}   
\def\physscr{\ref@jnl{Phys.~Scr}}   
\def\planss{\ref@jnl{Planet.~Space~Sci.}}   
\def\procspie{\ref@jnl{Proc.~SPIE}}   

\makeatother

\begin{document}

\begin{frontmatter}



\title{Intergalactic magnetic field lower limits up to the redshift $z\approx3$}


\author{Ievgen Vovk}

\affiliation{
    organization={Institute for Cosmic Ray Research, The University of Tokyo},
    addressline={5-1-5 Kashiwa-no-Ha}, 
    city={Kashiwa},
    postcode={277-8582}, 
    state={Chiba},
    country={Japan}
}

\begin{abstract}
Large-scale intergalactic magnetic fields may contain a mixture of galactic and cosmogenic contributions, that can be probed via observations of $\gamma$-ray ``echo'' -- a delayed emission from electromagnetic cascades initiated by the highest energy photons from the sources at cosmological distances. While these fields contributions may be disentangled based on the difference in their redshift evolution, thus far indications of non-negligible magnetic field have been found only at low redshifts.
This work aims at extending the intergalactic magnetic field constraints to redshifts $z\gtrsim 1$ using 17-year long all-sky observations of high-redshift active galactic nuclei with Fermi/LAT $\gamma$-ray telescope. Combing the Fermi/LAT measurements in the 0.1~GeV -- 1~TeV energy range with the Monte Carlo simulations of the $\gamma$-ray ``echo'', it is shown that the zero field strength hypothesis at $z = [0.5; 3]$ is disfavoured at the $\approx 8.6\sigma$ significance level, yielding the lower limit of $B \gtrsim 1\times10^{-18}$~cG for the magnetic field correlation length above 1~Mpc.
It is further shown that the same data put a lower limit on the volume-filling fraction of this field of $f\gtrsim 90\%$ in the redshift range considered.
It is also demonstrated that the derived limits are not substantially affected by either source flux variability or the assumed $\gamma$-ray emission attenuation model.
Implications of these limits for intergalactic magnetic field origin are discussed.
\end{abstract}

%


\end{frontmatter}


\section{Introduction}

Most of the known galaxies and galaxy clusters are believed to contain magnetic fields generated at the early stages of the their evolution or inherited from the intergalactic space during the proto-galaxy formation~\citep{kronberg94,grasso01,widrow02}. These initial ``seed'' fields could have been amplified by various dynamo effects during the galaxy lifetime and further ejected into the intergalactic space along with the supernovae or active galactic nuclei (AGN) driven outflows at redshifts $z\sim 1-4$~\citep{kulsrud08,beck08,marinacci18,garaldi21}. While the exact volume filling factor of such outflows is uncertain, it is plausible that intergalactic space may presently contain a mixture of cosmogenic and galactic magnetic fields.

First evidences for the presence of weak intergalactic magnetic fields (IGMF) were found in $\gamma$-ray observations of distant active galactic nuclei~\citep{Neronov10}.
On its way to Earth, the very-high-energy (VHE; $\gtrsim100$~GeV) emission from these objects is subject to partial absorption in interactions with the cosmic microwave background as well as infrared and optical photon fields constituting the extragalactic background light (EBL)~\cite{gould66, jelley66, stecker92}. The power absorbed is transferred to the electromagnetic cascades, initiated in the process, and eventually re-emitted as a lower-energy gamma-ray ``echo'', whose properties are sensitive to IGMF it develops within~\citep{aharonian_coppi,plaga95,neronov07,NeronovSemikoz09}. Lack of this ``echo'' signal in the observed AGN and gamma-ray burst (GRB) emission has been used to derive the lower limit on the intervening IGMF~\citep{Neronov10,Tavecchio11,dermer11,Taylor11,Vovk12, 2023A&A...670A.145A, 2023ApJ...950L..16A,Dzhatdoev:2023opo, Huang:2023uhw, 2024A&A...683A..25V}.
Coupled with the upper limits from cosmic microwave background fluctuations and AGN radio emission Faradey rotation~\citep[][and references therein]{NeronovSemikoz09, durrer13, 2019ApJ...878...92V, 2022MNRAS.515..256P}, these constrain IGMF strength to the $B\sim 10^{-17} - 10^{-11}$~G range (for field correlation length scale in excess of a megaparsec), but so far have been insufficient to identify its nature as both cosmogenic and galactic IGMF remain consistent with the data.

Even if the present-day IGMF constrains are inconclusive, the field nature may be probed via measurements at redshifts above $z>1$, owing to the distinctly different redshift evolution of the cosmogenic and galactic fields~\citep{durrer13, garaldi21, vazza25}. However, the increasing attenuation of the intrinsic VHE emission with redshift, at $z\simeq1$ reaching optical depth of unity already at 100~GeV energy~\citep{franceschini08}, makes these challenging from the observational point of view. With this, only a handful of VHE-band AGN detections have been reported thus far at $z>0.5$ in dedicated observational campaigns with ground-based $\gamma$-ray telescopes~\citep[e.g.][]{2008Sci...320.1752M, 2021A&A...647A.163M, 2015ApJ...815L..23A, 2015ApJ...815L..22A, 2016A&A...595A..98A}.
Furthermore, sparsity of such measurements in time results in them generally not being contemporaneous with the lower-energy ``echo'' constraints obtained with other instruments, dictating artificial assumptions on the AGN (non-)variability during the observational time window~\citep[e.g.][]{dermer11,Taylor11,Vovk12,2018ApJS..237...32A,2023A&A...670A.145A,2023ApJ...950L..16A}.



An opportunity to bypass these restrictions is offered by a deep, 17-years long all-sky exposure accumulated with Fermi/LAT $\gamma$-ray telescope~\citep{FermiLAT}, leading to a dozen of AGN detections in the VHE band at $z > 0.5$~\citep{4FGL,4LAC-DR3,2015A&A...575A..21N,2025arXiv250608497N}. In this manuscript, these detections are considered to search for the indications of IGMF-perturbed ``echo'' signal in the fully contemporaneous spectral energy distributions in the broad 0.1~GeV -- 1~TeV energy band. Applying the gamma-ray propagation code \CRBeam~\citep{CRBeam} to accurately describe the spectral and temporal properties of the ``echo'', it is shown that these observations alone yield IGMF lower limits up to a redshift of $z \approx 3$.

\section{Methods}

\subsection{High-z AGN sample}

The source sample used here was selected out of the known gamma-ray loud AGNs listed in the Fourth LAT AGN catalogue~\citep[4LAC-DR3;][]{4LAC-DR3} with the focus on high-redshift ($z>0.5$) sources that may present detectable ``echo''.
First and foremost, lower limits on IGMF with $\gamma$-ray observations require an ``echo'' signal to be distinguishable on top of the intrinsic source emission in case of a negligibly weak magnetic field. Such a condition is more easily satisfied for hard-spectrum AGNs, where the attenuated VHE emission power may exceed that intrinsically emitted at the ``echo'' energies~\citep{NeronovSemikoz09,Neronov10}, leading to a possibly dominant ``echo'' signal.
With this in mind, only a subset of 4LAC-DR3 AGNs with power law spectra indices\footnote{the choice of the spectral index from a power law fit here is made for the selection purposes only; a number of selected here AGNs feature curved spectral energy distributions. Still, the accepted sources with sufficient intrinsic emission curvature above few tens of GeV do not generate strong ``echo'' and thus do not impact the IGMF search performed here.} $\Gamma < 2$ (within the quoted uncertainties) was selected for the analysis here.
To increase the chances of detecting the ``echo'' signal, several alternative selection criteria were used to further narrow down the list. Firstly, all sources with the confident VHE signal (4LAC-DR3 detection test statistics in the VHE band $TS>25$) were selected. To include suitable sources where VHE signal could have been attenuated below Fermi/LAT detection level, bright AGNs with the reported flux in the adjacent 30-100~GeV energy band $\nu F_\nu^{30-100} > 1~\mathrm{eV / (cm^2~s)}$ were also added to the list. In addition to those, sources with probable VHE signal ($TS>9$) were further included. To include the sources whose VHE band detections could have been overlooked in 4LAC-DR3, VHE-detected high-redshift sources from Ref.~\cite{2025arXiv250608497N} were also considered. 
Finally, the subset of 4LAC-DR3 sources satisfying the above spectral index and flux requirements, where the lacking redshift information could have been recovered from SIMBAD astronomical database\footnote{\url{https://simbad.u-strasbg.fr/simbad/}}, was also included.
The resulting list of sources is given in Tab.~\ref{tab:high-z-agns}.

\begin{table*}
\centering
\caption{
    Hard-spectrum AGNs at $z>0.5$ selected here for analysis. 
    Columns list the 4FGL catalogue name, association, redshift,
    spectrum type (with ``PWL'' standing for power law and ``LP'' for log-parabola correspondingly), 
    power law spectral index, measured flux in the 30-100~GeV band, 
    detection test statistics (TS) square root (representing the detection significance)
    in the VHE band as 
    reported in 4LAC-DR3~\citep{4LAC-DR3} and 4FGL~\citep{4FGL} catalogues
    and the redshift value reference.
   The table is split into five sections: 
    (1) sources with confident VHE detection ($\sqrt{TS} > 5$),
    (2) bright sources with the flux in the adjacent 30-100~GeV energy band $\nu F_\nu^{30-100} > 1~\mathrm{eV / (cm^2~s)}$,
    (3) sources with possible VHE signal ($\sqrt{TS} > 3$),
    (4) VHE-detected high-redshift sources from Ref.~\cite{2025arXiv250608497N}
    and 
    (5) bright 4LAC-DR3 sources with spectroscopic redshifts recovered from Refs.~\cite{2019MNRAS.482.3458M, 2018AA...613A..51P, 2009ApJS..182..543A}.
    }
\label{tab:high-z-agns}

\begin{tabular}{l l c c c c c c}
\hline
4FGL name & Common name & z & Spec. & $\Gamma$ & $\nu F_\nu^{30-100}$ & $\sqrt{TS_{VHE}}$ & Ref.\\
\hline
\hline
4FGL J0033.5-1921 & KUV 00311-1938 & 0.61 & LP & 1.75 ± 0.02 & 3.9 ± 0.5 & 6.6 & \cite{4LAC-DR3} \\
4FGL J1037.7+5711 & GB6 J1037+5711 & 0.83 & LP & 1.89 ± 0.01 & 2.5 ± 0.4 & 10.1 & \cite{4LAC-DR3} \\
4FGL J1243.2+3627 & Ton 116 & 1.07 & LP & 1.77 ± 0.02 & 4.0 ± 0.5 & 6.4 & \cite{4LAC-DR3} \\
4FGL J1427.0+2348 & PKS 1424+240 & 0.604 & LP & 1.82 ± 0.01 & 12.5 ± 0.9 & 18.3 & \cite{4LAC-DR3} \\
\hline
4FGL J0035.2+1514 & RX J0035.2+1515 & 1.09 & LP & 1.83 ± 0.04 & 1.3 ± 0.3 & 2.0 & \cite{4LAC-DR3} \\
4FGL J0043.8+3425 & GB6 J0043+3426 & 0.966 & LP & 1.94 ± 0.02 & 2.2 ± 0.4 & 3.2 & \cite{4LAC-DR3} \\
4FGL J0630.9-2406 & TXS 0628-240 & 1.24 & LP & 1.81 ± 0.02 & 6.4 ± 0.7 & 1.7 & \cite{4LAC-DR3} \\
4FGL J1517.7+6525 & 1H 1515+660 & 0.702 & LP & 1.83 ± 0.03 & 1.4 ± 0.3 & 2.3 & \cite{4LAC-DR3} \\
4FGL J2357.4-1718 & RBS 2066 & 0.85 & PWL & 1.72 ± 0.07 & 1.2 ± 0.3 & 2.3 & \cite{4LAC-DR3} \\
\hline
4FGL J0649.5-3139 & NVSS J064933-313917 & 0.563 & LP & 1.68 ± 0.08 & 0.6 ± 0.2 & 3.9 & \cite{4LAC-DR3} \\
4FGL J1101.4+4108 & RX J1101.3+4108 & 0.58 & PWL & 1.87 ± 0.08 & 0.5 ± 0.2 & 3.1 & \cite{4LAC-DR3} \\
4FGL J1309.4+4305 & B3 1307+433 & 0.691 & LP & 1.91 ± 0.02 & 0.9 ± 0.2 & 4.5 & \cite{4LAC-DR3} \\
4FGL J1340.5+4409 & RX J1340.4+4410 & 0.546 & PWL & 1.74 ± 0.12 & 0.3 ± 0.2 & 4.6 & \cite{4LAC-DR3} \\
\hline
4FGL J0045.3+2128 & GB6 J0045+2127 & 2.07 & LP & 1.81 ± 0.03 & 2.4 ± 0.4 & 3.8 & \cite{2020ApJS..250....8L} \\  
4FGL J0915.9+2933 & Ton 0396 & 1.52 & LP & 1.88 ± 0.02 & 2.0 ± 0.4 & 4.3 & \cite{2018AA...618A..80G} \\  
4FGL J1425.0+3615 & FBQS J142455.5+361536 & 1.09 & LP & 1.88 ± 0.05 & 0.8 ± 0.2 & 0.0 & \cite{2009ApJS..182..543A} \\  
4FGL J1448.0+3608 & RBS 1432 & 1.51 & LP & 1.82 ± 0.02 & 1.7 ± 0.3 & 7.1 & \cite{2009ApJS..182..543A} \\ 
\hline
4FGL J0120.4-2701 & PKS 0118-272 & 0.557 & LP & 1.92 ± 0.02 & 3.3 ± 0.5 & 6.5 & \cite{2019MNRAS.482.3458M} \\    
4FGL J1118.0+5356 & RX J1117.9+5356 & 2.95 & LP & 1.88 ± 0.03 & 1.1 ± 0.2 & 2.4 & \cite{2018AA...613A..51P} \\  
4FGL J1248.3+5820 & PG 1246+586 & 0.847 & LP & 1.91 ± 0.01 & 5.4 ± 0.5 & 7.8 & \cite{2009ApJS..182..543A} \\  
\hline
\end{tabular}
\end{table*}

For each of the selected sources, a time-averaged spectral energy distribution was extracted from the Fermi/LAT data over the August~2008 to July~2025 period using the \Fermitools\ package v2.2.0 and \FermiPy\ framework v1.2.0~\citep{fermipy}, as described in the \Fermitools\footnote{\url{https://fermi.gsfc.nasa.gov/ssc/data/analysis/}} and \FermiPy\footnote{\url{https://fermipy.readthedocs.io/}} documentation. The event selection comprised $\gamma$-ray photons corresponding to the {\tt P8R3~SOURCE} type, recorded with $10^\circ$ from the catalogue source positions in the 0.1~GeV -- 1~TeV energy range. Only events satisfying the {\tt (DATA\_QUAL>0) \&\& (LAT\_CONFIG==1)} good time intervals criteria were retained in the analysis. The data reduction accounted for both galactic and extragalactic diffuse emission (parametrized in the {\tt gll\_iem\_v07.fits} and {\tt iso\_P8R3\_SOURCE\_V2\_v1.txt} files provided with \Fermitools) as well as the known sources from the LAT 14-year Source Catalog~\cite[4FGL-DR4;][]{4FGL,4FGL-DR4} within $17.5^\circ$ from the source position with the spectral shapes taken from the catalogue. Throughout the spectrum extraction procedure, only the sources within $10^\circ$ from the source position were left free and only those within $5^\circ$ were allowed to optimize the model spectral shape in addition to its normalization. To improve the analysis sensitivity, as well as reduce the systematical uncertainties and the source confusion, the likelihood components were constructed separately for each of the Fermi/LAT PSF event types~\citep{FermiLAT_Pass7_validation}, restricting the usage of the worst event type quartile (``PSF0'' corresponding to {\tt evtype=4}) to energies greater than 1~GeV and the second to worst (``PSF1'' corresponding to {\tt evtype=8}) -- to those above 0.3~GeV. All event types were used jointly in the fit, that also accounted for the energy dispersion for all components save the extragalactic diffuse one.

\subsection{``Echo'' emission calculation}

For each of the sources selected and processed above, an individual ``echo'' model was constructed using the \CRBeam\ cosmic ray propagation code, shown to adequately handle ``echo'' development in the presence of IGMF at least up to the redshift of $z\simeq 1$~\citep{CRBeam}. The code was set up to inject $10^6$ photons in the energy range between 0.1~GeV and 1~TeV at the individual redshifts of the sources from Tab.~\ref{tab:high-z-agns} and propagate them towards the observer through the intergalactic space filled with IGMF with the correlation length of $\lambda_B=1$~Mpc and the physical (related to the comoving as $B_{phys} = (1+z)^2 B_{comoving}$) strength ranging\footnote{
    For $\lambda_B=1$~Mpc, magnetic fields below $B_{lim} \simeq 10^{-21}$~G result in time delays smaller than that intrinsic to the ``echo'' in the entire energy range considered and thus, in principle, can not be detected~\citep{NeronovSemikoz09}.
}
from $10^{-21}$~G to $3\times10^{-16}$~G
-- as well as cosmic microwave background and low-energy photon fields following Ref.~\cite{franceschini08}. Due to the redshift limitation of the latter, for sources at $z>2$ the Ref.~\cite{stecker16} model at its lowest photon density level within the quoted uncertainties was used instead (comparison of different EBL models is presented in~\ref{sect:ebl-comparison}).
Upon the completion of the photon propagation, the respective time delay of the ``echo'' emission was evaluated from the source-to-observer and the final source-to-photon distances difference; these calculations were found to be in good agreement with the earlier estimates from Ref.~\citep{Taylor11}.

Performed \CRBeam\ calculations on their own correspond to instantaneous photon injection in an infinitesimal beam opening at the source and require post-processing to match with the Fermi/LAT data. First, the recorded photons were uniformly scattered across the cone with $1^\circ$ opening, mimicking the AGN jet.
Temporal distribution of the photons, matching the data, was then obtained convolving the recorded arrival times with the step function product $\theta(t-t_{start}) \times \theta(t_{stop} - t)$ with $t_{start}$ and $t_{stop}$ representing the beginning and the end of Fermi/LAT observations time range. Lastly, the recorded events were re-weighted according to the ratio of the assumed and injected emission models and binned to match the considered Fermi/LAT energy intervals.

The source emission model obtained this way self-consistently contains the intrinsic and ``echo'' components. The model is fully parametrized by the assumed intrinsic spectral shape and can be directly fit to the data. To this end the predicted model flux in each energy bin was compared against the measurements interpolating the ``flux -- likelihood'' profile automatically computed by \FermiPy\ and stored along with the extracted spectral data. Likelihood maximization itself was performed with the \iminuit\ package~\cite{iminuit}. The systematical uncertainties on the Fermi/LAT collection area were propagated to the fit by stretching the likelihood in each bin around its maximum by 3\% below 100~GeV and $3\% + 12\% \times (\log(E/\mathrm{MeV}) - 5)$ above this energy\footnote{\url{https://fermi.gsfc.nasa.gov/ssc/data/analysis/scitools/Aeff_Systematics.html}}.
As each energy bin is treated independently, this procedure, in principle, allows for larger systematical model-to-data deviations compared to the bracketing approach suggested for Fermi/LAT data analysis~\citep{FermiLAT_Pass7_validation}, conservatively reducing the analysis sensitivity to the putative ``echo'' signal.

\section{Results}

Before attempting to derive the IGMF constraints, the modelling assumptions put forward in the previous section were verified.
Spatial extension of ``echo'' emission was estimated with \CRBeam\ simulations and for time delays below 17~yr, corresponding to the Fermi/LAT observations duration, was found to be less than $10^{-3}$~deg in the entire IGMF strength range considered (see~\ref{sect:echo-extension-tdelay}). As this is well below the Fermi/LAT PSF 68\% containment, the assumed $1^\circ$ emission cone opening and point-like treatment of the sources in Fermi/LAT analysis do not introduce any tangible uncertainty to the derived fluxes.
At the same time, the constant in time ``echo'' flux injection assumption may not necessarily work in case of source variability within the Fermi/LAT observational time window 
-- due to the difference in the fluence integration between the intrinsic and time-delayed ``echo'' emission.
However, for the sources from Tab.~\ref{tab:high-z-agns} the cumulative fluence variations with respect to the mean flux do not exceed 5-10\% -- with exception of RX~J1340.4+4410, for which ``echo'' prediction based on the constant flux assumption may introduce a sizeable systematical uncertainty (see~\ref{sect:variability}). Still, as is detailed below, contribution of this object to the IGMF constraint is not decisive.

Spectral energy distributions (SED) of the sources from Tab.~\ref{tab:high-z-agns}, derived here, are consistent with the 4FGL records~\citep{4FGL}; no new significant detections in the VHE range, except those already reported in 4FGL and Ref.~\citep{2025arXiv250608497N}, have been found.
Accounting for the VHE emission attenuation, derived SEDs generally match the phenomenological power law and log-parabola shapes indicated in the catalogue.
The spectral models indicated in 4LAC-DR3 have been thus taken as the proxy for the intrinsic spectra\footnote{
    In case of Ton~116, phenomenological log-parabola and exponentially cut off power law models are indistinguishable with the best-fit likelihood value difference $\Delta \log L \approx 0.06$.
    Though the reconstructed SED indicates no curvature below VHE energies, favouring the power law interpretation, the log-parabola model, featuring an additional curvature-related parameter, was kept here as it conservatively results in a weaker ``echo'' for a given IGMF strength.
}; however, to conservatively restrict the total model emitted power in the VHE band for the spectra with little intrinsic curvature, an additional free exponential cut off has been introduced in all cases.

\subsection{IGMF strength lower limit}

\begin{figure*}
    \includegraphics[width=0.5\linewidth]{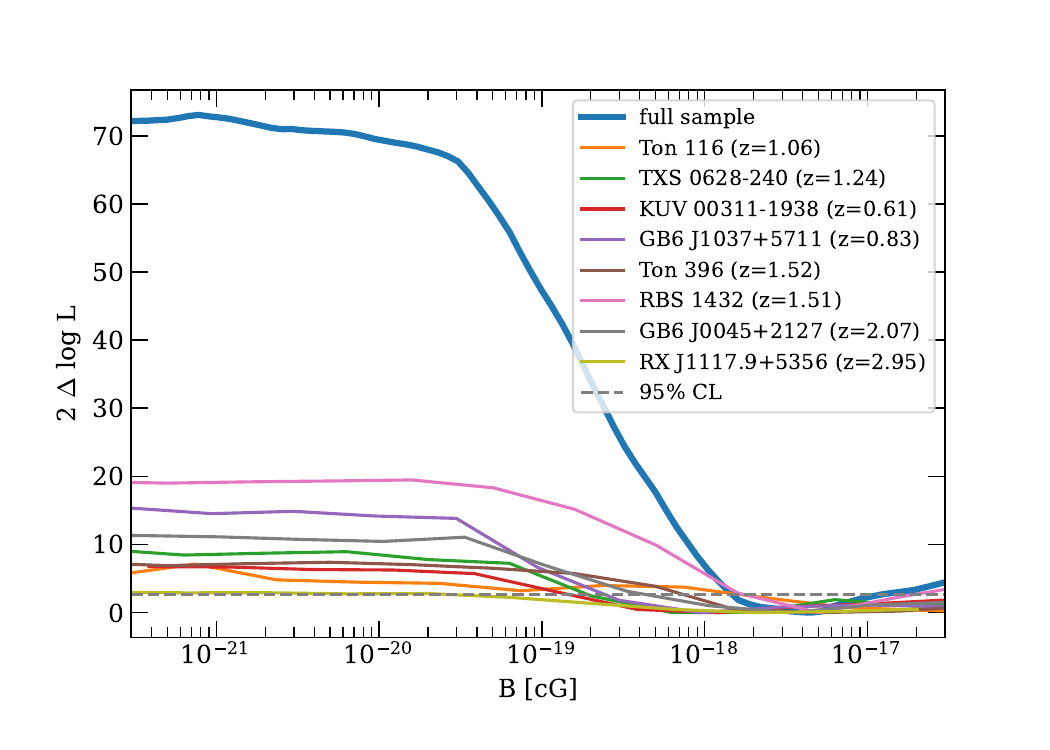}
    \includegraphics[width=0.5\linewidth]{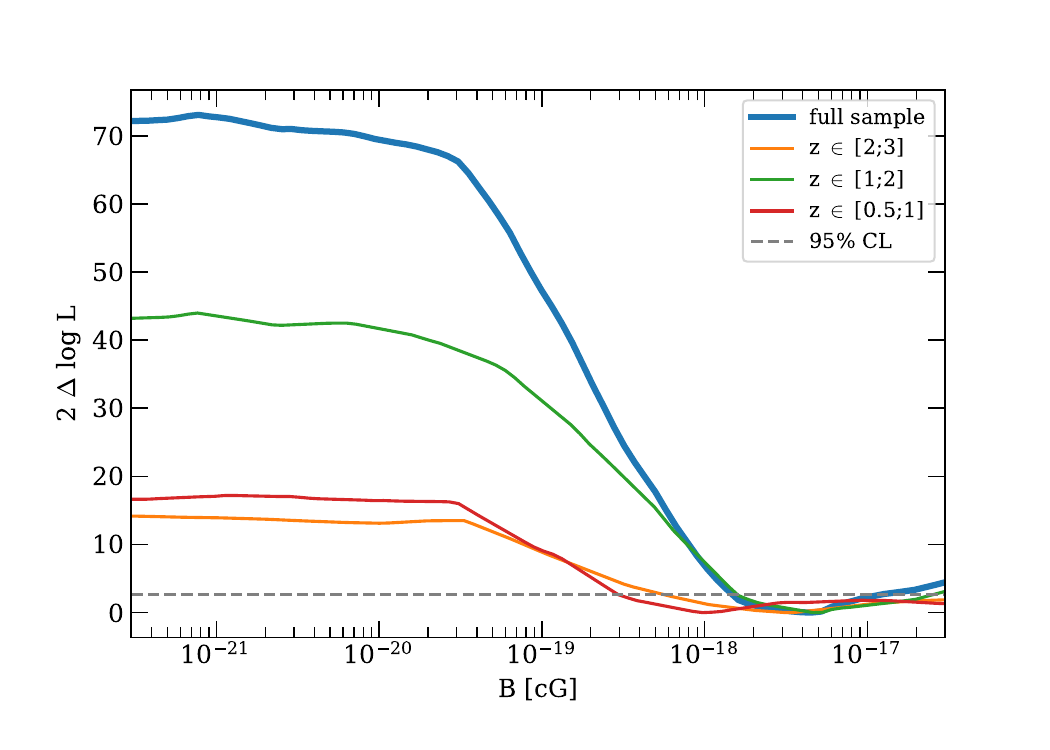}
    \caption{
        Best-fit likelihood value difference with respect to its minimum as a function of the assumed IGMF comoving field strength.
        \textit{Left:} contributions of sources that produce individual field constraints at more than 95\% confidence level.
        \textit{Right:} contributions of the sources in different redshift bins.
        Horizontal dashed line in both panels represents the threshold value $2\Delta \log L \approx 2.71$ corresponding to the one-sided 95\% confidence level lower limit.
    }
    \label{fig:igmf-loglike}
\end{figure*}

The obtained best-fit likelihood profile as a function of IGMF strength is given in Fig.~\ref{fig:igmf-loglike}. With all sources combined, the zero IGMF hypothesis is rejected at approximately $8.6\sigma$ significance level.
At 95\% confidence level, the resulting sample-averaged IGMF limit at $z>0.5$ can be summarised as
\begin{equation}
    B \gtrsim \left\{
    \begin{array}{lr}
        1 \times 10^{-18}~\mathrm{cG} &, \lambda_B \ge D_e \\
        1 \times 10^{-18} \left( \lambda_B / D_e \right)^{-1/2}~\mathrm{cG} &, \lambda_B < D_e \\
    \end{array}
    \right.
    \label{eq::igmf_limit}
\end{equation}
where the transition value of the comoving IGMF correlation length $\lambda_B$ is set by the electron cooling distance on inverse Compton scattering of CMB photons~\cite{NeronovSemikoz09}
\begin{equation}
    \begin{split}
        D_e(z) = (1+z) \frac{3 m_e c^4}{4 \sigma_T U_{CMB}^\prime E_e^\prime} \\
        = \frac{\sqrt{3}}{8} (1+z)^{-3} \frac{m_e c^3}{\sigma_T \sigma_{SB} T_{CMB}^4}
            \sqrt{\frac{\epsilon_{CMB}}{E_\gamma}} \\
        \approx 2 (1+z)^{-3} \sqrt{\frac{0.1~\mathrm{GeV}}{E_\gamma}}~\mathrm{cMpc}
    \end{split}
\end{equation}
with $E_\gamma$ representing the observed ``echo'' emission energy, $U_{CMB}$, $T_{CMB}$ and $\epsilon_{CMB}$ denoting the cosmic microwave background energy density, temperature and photon energy, $\sigma_T$ being the Thomson scattering cross-section, $\sigma_{SB}$ -- Stefan-Boltzmann constant and primed values corresponding to the source reference frame.
As shown in the right panel of Fig.~\ref{fig:igmf-loglike}, the obtained $B$ limit value is comparable to that corresponding to the $z \in [1;2]$ sub-sample and 2-3~times the individual limits derived in the $z \in [1;2]$ and $z \in [2;3]$ redshift bins.
The larger in comparison limit value in the $z \in [1;2]$ bin is set by contributions of RBS~1432 and Ton~116, both displaying little curvature in the measured spectra along with the highly significant detections in the VHE band. They thus require a stronger IGMF to suppress the ``echo'' signal extending to higher (relative to the other sources) energies. Without these two sources, the $z \in [1;2]$ limit becomes comparable to that obtained in other redshift bins.

\begin{figure*}
    \includegraphics[width=1.0\linewidth]{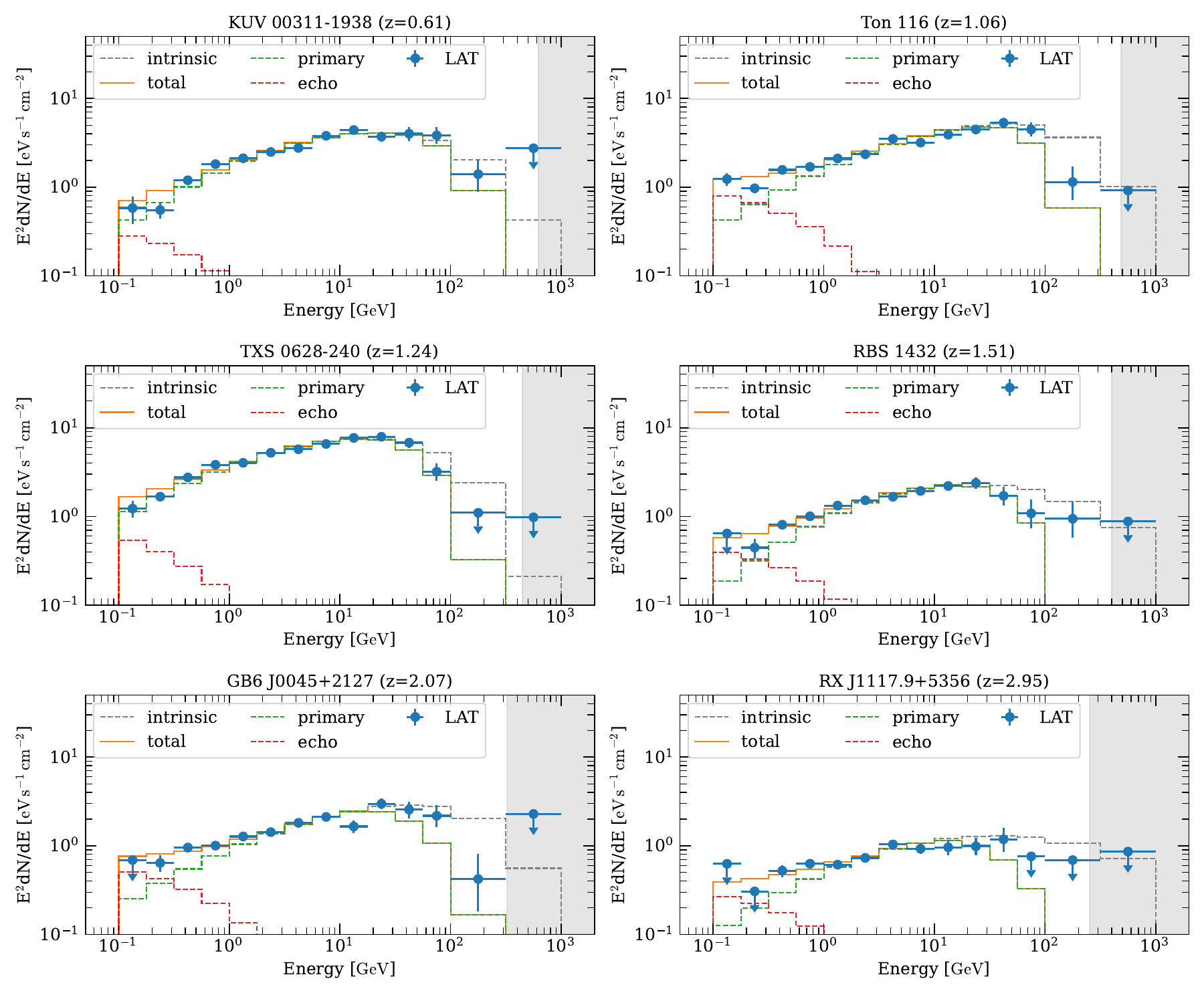}
    \caption{
        Best-fit ``echo'' spectral energy distribution models for the representative sources with the strongest individual contribution to the derived IGMF lower limit at their redshifts, listed in Tab.~\ref{tab:results}. Data points represent the Fermi/LAT measurements obtained here; flux uncertainties corresponding to 68\% confidence level (CL) whereas the upper limits, plotted where the detection test-statistics is $TS < 10$, were calculated for 95\% CL. Solid orange lines depict the best-fit spectral model constituting the sum of the source primary emission (dashed green lines) and the IGMF-modified ``echo'' (dashed red lines), evaluated in the energy bins of Fermi/LAT measurements. Dashed grey lines represent the intrinsic primary emission before attenuation on the extragalactic photon fields. Grey shaded region marks the energies above the maximal simulated $1~\mathrm{TeV} / (1+z)$ energy. All models are evaluated for the weakest considered IGMF with $B=10^{-21}$~cG and $\lambda_B = 1$~cMpc.
    }
    \label{fig:best-fit-sed}
\end{figure*}

However, not all the sources considered contribute to the derived limit in the same way. Several sources listed in Tab.~\ref{tab:results} enable IGMF lower limits to be derived at the confidence level above 95\% on the source-by-source basis.
The strongest field limits are set by RBS~1432, Ton~116 and Ton~396, as one can also see from Fig.~\ref{fig:igmf-loglike}.
Several sources though present the trend for the increasing best-fit likelihood value for IGMF above $B \sim 10^{-17}$~cG, suggesting a two-sided field strength measurement of $B \in [10^{-18}; 1\times10^{-17}]$~cG at 95\% confidence level. However, the significance of the corresponding {\it upper} limit, dominated by RBS~1432, KUV~00311-1938 and GB6~J0045+2127, does not exceed $2.8\sigma$.
If these sources are excluded from the analysis, a one-sided IGMF lower limit $B \gtrsim 1\times10^{-18}$~cG is still warranted by the rest of the sources at approximately $6\sigma$ significance level.
The individual estimates of IGMF strength for the considered sources are listed in Tab.~\ref{tab:results}. For representative sources with the strongest individual contribution to the derived IGMF lower limit the spectral energy distributions and best-fit models for the minimal considered IGMF strength $B=10^{-21}$~G for are shown in Fig.~\ref{fig:best-fit-sed}.

\begin{table*}
\centering
\caption{
    IGMF limits derived from individual sources in Tab.1, that exclude the minimal considered 
    $B_{phys}=10^{-21}$~G magnetic field at least at 95\% confidence level.
    The quoted one-sided field lower limits $\log(B_{95})$ correspond to 95\% confidence level,
    while $S_{IGMF}$ represents the $B_{min}$ exclusion significance in gaussian sigma units; 
    $\sqrt{TS_{100-300}}$ denotes the source detection significance in the 100-300~GeV energy band.
    For sources that indicate a worsening model fit for stronger fields, $\log(B)$ values indicate
    the best-fit value and the corresponding 68\% confidence range for the IGMF strength.
    }
\label{tab:results}
\begin{tabular}{l r c c c c}
\hline
Common name & z & $\sqrt{TS_{100-300}}$ & $S_{IGMF}$ & $\log(B_{95})$ & $\log(B)$ \\
 &  &  & [$\sigma$] & [dex(cG)] & [dex(cG)] \\
\hline
\hline
GB6 J0045+2127 & 2.07 & 3.2 & 3.6 & -17.4 & -16.5 ± 0.5 \\
GB6 J1037+5711 & 0.83 & 9.5 & 4.1 & -18.1 & -16.8 ± 0.9 \\
GB6 J1424+3615 & 1.09 & 3.3 & 2.1 & -18.5 &  --\\
KUV 00311-1938 & 0.61 & 6.3 & 2.8 & -18.4 & -17.5 ± 0.6 \\
RBS 1432 & 1.51 & 5.9 & 4.5 & -16.9 & -16.5 ± 0.2 \\
RX J1117.9+5356 & 2.95 & 2.0 & 2.0 & -18.4 &  --\\
TXS 0628-240 & 1.24 & 2.7 & 3.3 & -18.0 & -16.8 ± 0.8 \\
Ton 116 & 1.06 & 6.3 & 2.6 & -17.1 &  --\\
Ton 396 & 1.52 & 4.9 & 2.9 & -17.3 & -16.3 ± 0.8 \\
\hline
\end{tabular}
\end{table*}

%


The above IGMF lower limits do not critically depend on the choice of the maximal photon energy $E_{max}=1$~TeV injected in the \CRBeam\ simulation.
Corrected for redshift, $E_{max}$ is close to the maximal energy at which the sources with the strongest contribution are detected, as one can see in Fig.~\ref{fig:best-fit-sed}.
Furthermore, the ubiquitously introduced high-energy exponential cut-off and intrinsic spectral curvature result in soft intrinsic source spectra above $\sim 100$~GeV, reducing the attenuated power up to and above $E_{max}$.
This implies that further extension of $E_{max}$ far beyond 1~TeV energy also would not have a substantial effect on the predicted ``echo'' signal and thus the derived limits.


The choice of the VHE attenuation model, at the same time, directly affects the IGMF limits. However, the Ref.~\cite{franceschini08} model, employed here, remains a conservative choice up to $z=2$, predicting roughly 20-60\% less of total VHE attenuated power at redshift $z\approx1$ compared to the other models supported by \CRBeam . Same can be said about the most recent attenuation models from Refs.~\cite{saldana21, finke22}, with the former predicting $7-15\%$ less attenuation depending on the redshift only at its lower uncertainty level. At redshifts $z>2$ the employed Ref.~\cite{stecker16} model suggests attenuation similar to that of Ref.~\cite{saldana21} with other available in \CRBeam\ models predicting $13-80\%$ larger value.
The performed above calculations may thus overestimate the ``echo'' power by not more than 20\% -- insufficient to explain the lack of the ``echo'' signal under the zero IGMF hypothesis. Indeed, while forcing the correspondingly lower ``echo'' signal in the calculations reduces the zero IGMF exclusion significance to $\approx 7.6\sigma$, it lowers the numerical limit value quoted above only twofold.

\subsection{IGMF volume-filling fraction}

Performed simulations were also used to constrain the volume-filling fraction $f$ of IGMF, defined as a fraction of the simulated volume space occupied by the non-trivial field exceeding the weakest detectable strength $B_{lim} \simeq 10^{-21}$~G~\cite{NeronovSemikoz09}. To this end the source flux model $F(B, \theta_1 ... \theta_n)$, where $\theta_1 ... \theta_n$ represent the source intrinsic spectrum parameters, was replaced with $\widetilde{F}(B, f, \theta_1 ... \theta_n) = f F(B, \theta_1 ... \theta_n) + (1-f) F(B_{lim}, \theta_1 ... \theta_n)$. Varying the volume-filling fraction in the range $f \in [0.01;1]$ the balance in the model was shifted between the $B_{lim}$ and the target field value for each of the sources and IGMF strength values considered above. 

\begin{figure}
    \includegraphics[width=\columnwidth]{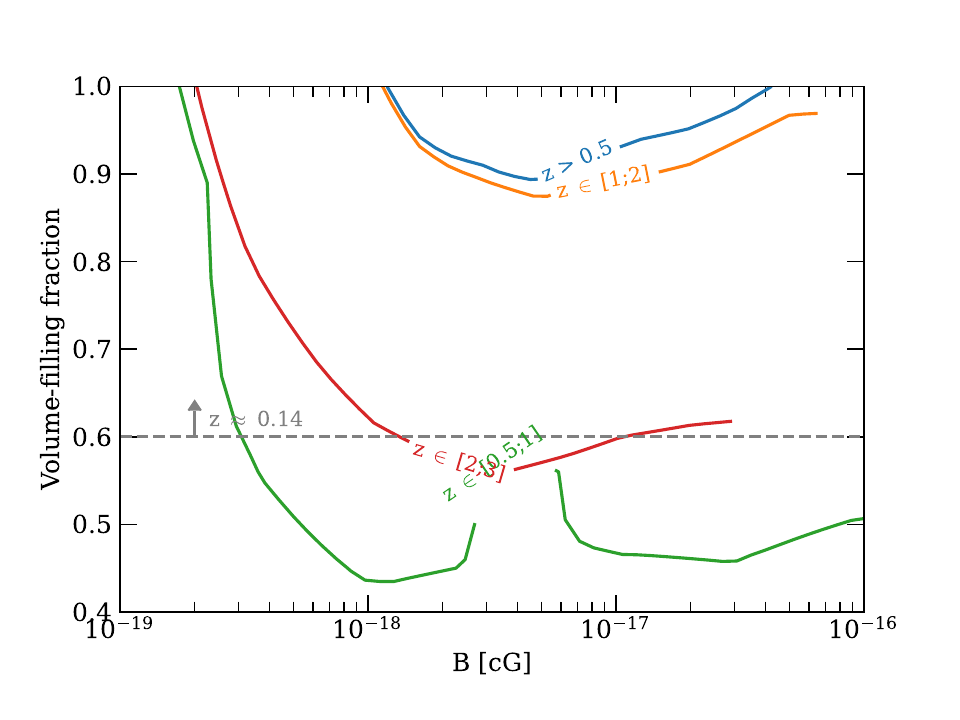}
    \caption{
        The derived 95\% confidence level lower limit on the combination IGMF strength and its volume-filling fraction. Different colors represent the limits derived in separate redshift bins as indicated in the figure; $z>0.5$ denotes the total limit. The dashed horizontal line represents the IGMF volume filling fraction constraint derived at the redshift $z\approx0.14$ in Ref.~\cite{dolag11}.
    }
    \label{fig:vff}
\end{figure}

The resulting limit on $f$ as a function of IGMF is given in Fig.~\ref{fig:vff}. Marginalized over the IGMF values, the minimal IGFM volume-filling fraction required for the model agreement with Fermi/LAT data is $f > 0.96$ at 95\% confidence level. A similarly strong constraint is obtained in the $z \in [1;2]$ redshift window, dominated by RBS~1432. Exclusion of this source relaxes the $z \in [1;2]$ limit to $f>0.87$, close to values derived in the other two redshift bins ($f>0.65$ and $f>0.78$ for $z \in [0.5;1]$ and $z \in [2;3]$ correspondingly).

\section{Discussion}

The derived here limits present the first attempt to constrain IGMF using single-instrument observations of both intrinsic and ``echo'' emission. Compared to earlier works, this approach eliminates the need in artificial assumptions regarding the source (non-)variability~\citep[e.g.][]{dermer11,Taylor11,Vovk12,2018ApJS..237...32A,2023ApJ...950L..16A} or spectral shape stability~\citep[e.g.][]{2023A&A...670A.145A}, representing the key practical uncertainties behind earlier IGMF searches. 

\begin{figure}
    \includegraphics[width=\columnwidth]{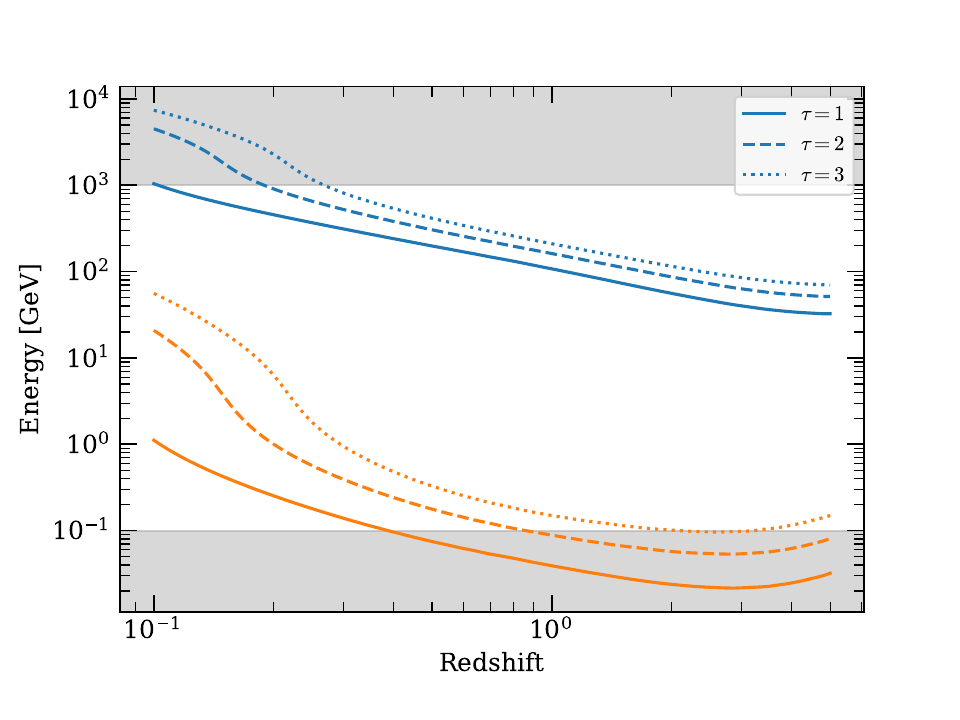}
    \caption{
        Source primary (blue) and characteristic ``echo'' (orange) photon energies as a function of redshift and $\gamma$-ray attenuation optical depth $\tau$ for EBL model from Ref.~\cite{saldana21}. Grey shaded regions mark the energies outside of the 0.1~GeV -- 1~TeV range considered here.
    }
    \label{fig:primary-secondary-energies}
\end{figure}

The single-instrument approach, however, is not universal as it requires that both the attenuated emission and ``echo'' are within the energy range of the instrument. An approximate condition for this can be evaluated analytically, assuming the ``echo'' originating from a primary energy $E_{\gamma,0}$ is not attenuated itself and in the source reference frame corresponds to the characteristic emission energy $E_\gamma^\prime = \frac{4}{3} \epsilon_{CMB}^\prime (E_e^\prime / m_e c^2)^2$ of electrons with $E_e^\prime = E_{\gamma,0}^\prime / 2$ up-scattering the CMB radiation with peak energy $\epsilon_{CMB}^\prime \approx 2.8 k_B T_{CMB}^\prime \approx 6.6\times10^{-4} (1+z)$~eV, yielding
\begin{equation}
    E_\gamma \approx 0.08 (1+z)^2 \left( \frac{E_{\gamma,0}}{0.3~\mathrm{TeV}} \right)^2~\mathrm{GeV}
\end{equation}
The corresponding primary and ``echo'' energies, evaluated at few representative optical depth, are shown in Fig.~\ref{fig:primary-secondary-energies}. It can be seen from there that while at redshifts $z>0.3$ the primary source emission up to the attenuation optical depth $\tau=3$ is fully contained within the Fermi/LAT energy range,
such measurements alone may be not suitable for nearby sources, where the majority of IGMF lower limits has been derived thus far. Likewise, higher-redshift Fermi/LAT measurements can constrain only the high-energy tail of the ``echo'' emission -- even if the attenuated flux would be accurately measured with the next-generation VHE facilities like CTA~\citep{cta}. To boost the ``echo'' sensitivity, those would have to be paired with dedicated observations with MeV-band instruments~\citep[such as AMEGO, e-ASTROGAM or MAST;][]{amego, e-astrogam, mast}.

It should be noted, that the limits obtained here may be affected by the development of the plasma instabilities, suggested to be capable of dissipating the absorbed emission power before it is re-emitted in the form of the ``echo''~\cite[e.g.][]{broderick12}. While presently there is no self-consistent description of this problem, recent studies indicate only a marginal role of the instabilities in reduction of the emission power available to the ``echo''~\cite{perry21,alawashra25}. As such, the effect of these instabilities was not considered here.

\begin{figure}
    \includegraphics[width=\columnwidth]{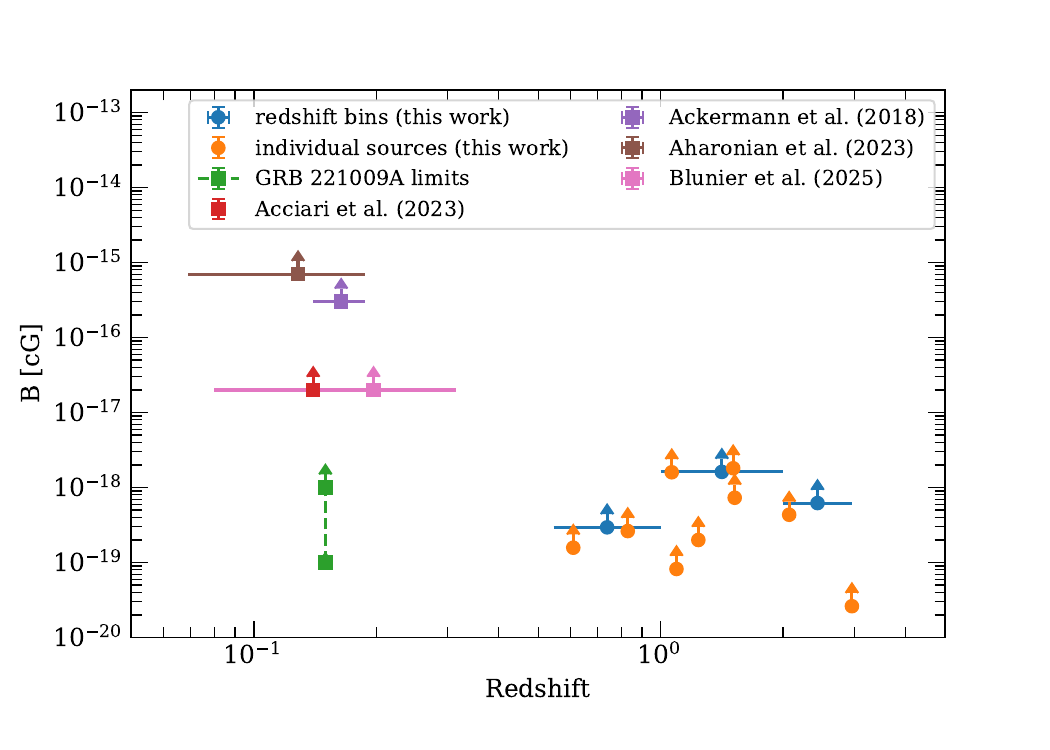}
    \caption{
        Comparison of the comoving IGMF strength constraints obtained here at 95\% confidence level for $\lambda_B=1$~Mpc -- both sample-averaged (blue) and for individual sources (orange) -- with those derived earlier at lower redshifts. For sample-averaged constraints from literature~\citep{2018ApJS..237...32A, 2023A&A...670A.145A, 2023ApJ...950L..16A, 2025arXiv250622285B} the uncertainties indicate the redshift range of the sources employed; in case of GRB~221009A, a range of limits from Refs.~\cite{Dzhatdoev:2023opo, Huang:2023uhw, 2024A&A...683A..25V} is shown.
    }
    \label{fig:igmf-vs-z}
\end{figure}

The IGMF limits derived here are lower than those obtained earlier from AGN observations, as shown in Fig.~\ref{fig:igmf-vs-z}.
In part this is due the choice of the VHE attenuation model, selected to minimize the predicted ``echo'' power for a given source emission spectrum. While this choice reduces the IGMF exclusion power in this work, it makes the derived limits robust against existing uncertainties of the VHE attenuation.
Furthermore, these limits provide the first direct evidence for the presence of IGMF at $z \gtrsim 1$, enabling tests of its cosmological evolution.

Mean free path of VHE photons with energy below 1~TeV remains $\lambda_{mfp} \gtrsim 100$~Mpc up to $z \simeq 3$, so the obtained limits probe the large-scale magnetic field filling the cosmological voids of the Universe. It should be noted that field strength probed here is not equivalent to the volume-averaged one as ``echo'' constraints are weakly sensitive to the magnetic field strengths beyond
\begin{equation}
    \begin{split}
        B_{iso} = 2\pi \frac{E_e}{D_e c} \\
            = 4 (1+z)^2 \frac{\sigma_T \sigma_{SB}}{c^2} T_{CMB}^4 \frac{E_\gamma}{\epsilon_{CMB}} \\
            \approx 6 \times 10^{-16} (1+z)^2 \frac{E_\gamma}{0.1~\mathrm{GeV}}
            \mathrm{cG}
    \end{split}
\end{equation}
which isotropise the ``echo'' emission so that its expected flux becomes almost field-independent. Nevertheless, as shown above, the IGMF probed here occupies a significant $f \gtrsim 0.8$ fraction of the intergalactic space volume at $z \gtrsim 1$.

The galactic origin of comparably strong magnetic field on these spatial scales is disputed by the modelling of magnetogenesis around young galaxies~\citep{dolag11,garaldi21,garcia21,vazza25}, suggesting the volume-filling fraction of $B>10^{-18}$~cG IGMF generated this way does not exceed $\sim 40\%$ even at $z\sim0$. With the latter decreasing quickly above $z\sim 1$~\citep{beck13,garaldi21,garcia21,vazza25}, the lower limits obtained here at $z\sim 2-3$ appear inconsistent with these calculations.
Small volume-filling fraction of galactic-origin fields at $z\gtrsim 1$ implies the presented limits probe IGMF largely unaffected by the galactic feedback, pointing to its primordial origin. In this context it is worthy to point out that stronger IGMF limits at redshifts $z\sim0$ do not necessarily constrain the same component of the present-day IGMF as the latter may still be partially contaminated by the galactic feedback.
In this respect, IGMF measurements at high and low redshifts
may be eventually used to differentiate the galactic and cosmological contributions to the present-day field.
This possibility may be further clarified with further observations with dedicated VHE instruments, such as LHAASO~\cite{2024ApJS..271...25C} and CTA~\citep{cta} at low and high redshifts respectively.




\appendix

\section{EBL models comparison}
\label{sect:ebl-comparison}

\begin{figure}
    \includegraphics[width=\columnwidth]{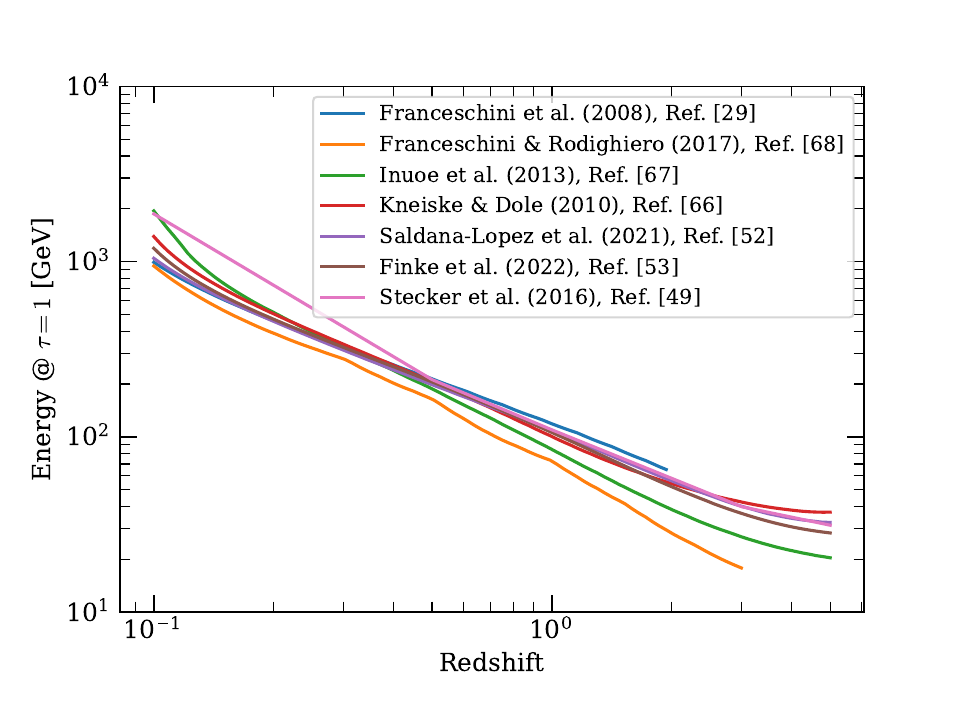}
    \caption{
        Comparison of the energy at the optical depth $\tau=1$ for EBL models~\cite{franceschini08, kneiske10, inoue13, franceschini17, saldana21, stecker16, finke22} supported in \CRBeam\ and the newer one from Ref.~\cite{saldana21}. Except for Ref.~\cite{stecker16} model (brown curve), the energies were calculated with the {\it ebltable} package~\cite{ebltable}.
        For models from Ref.~\cite{franceschini08} and \cite{franceschini17}, depicted by blue and orange lines, values shown correspond to the redshift range covered by the tables in the original publications.
    }
    \label{fig:ebl-comparison}
\end{figure}

Spectral energy density of EBL directly affects the attenuation of VHE emission and, consequently, the strength of the expected ``echo'' signal, proportional to the attenuated power. The transmission of the various EBL models, including those supported by \CRBeam, is shown in Fig.~\ref{fig:ebl-comparison} in terms of the $\gamma$-ray photon energy corresponding to the optical depth of $\tau=1$. It can be seen that in the redshift range $z>0.5$ considered here the Ref.~\cite{franceschini08} model consistently yields the highest $\gamma$-ray energy (i.e. a lower attenuation factor) up to the maximal $z=2$ where it is defined. In the redshift window $z \in [2;3]$ model in Ref.~\cite{stecker16}, included in \CRBeam, is similar to the more recent one from Ref.~\cite{saldana21}.

As not all EBL models are supported by \CRBeam, the possible impact of the model change was estimated integrating the total attenuated power in each case for the best-fit spectral parameters of the studies sources, yielding the values quoted in the main text.

\section{High-redshift ``echo'' signal spatial extension and time delay}
\label{sect:echo-extension-tdelay}

\begin{figure}
    \includegraphics[width=\columnwidth]{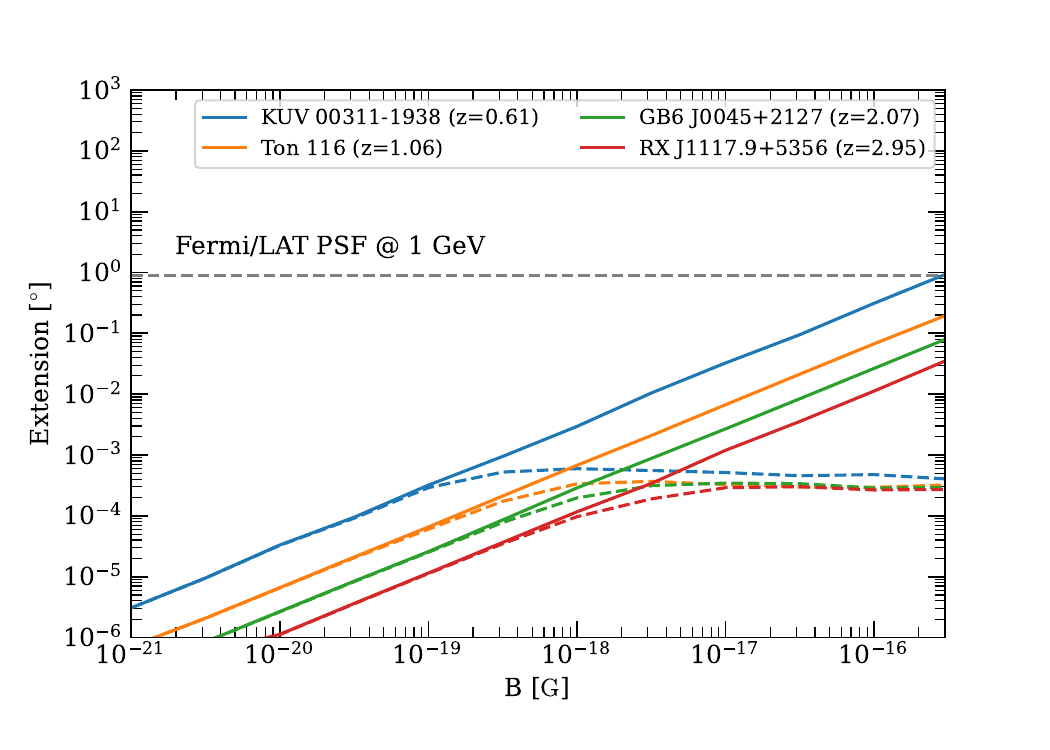}
    \caption{
        Spatial extension (68\% containment radius, solid lines) of the ``echo'' emission in the $E_\gamma=0.25-0.35$~GeV energy range, calculated for several representative sources from Tab.~\ref{tab:high-z-agns}, covering the broad redshift range $z\approx0.6-1.2$. Dashed lines of the same colors denote the ``echo'' extension if only time delays of $t_d < 17$~yr, corresponding to the duration of Fermi/LAT observations, are considered. For comparison, the grey dashed line marks the Fermi/LAT angular resolution (same 68\% containment, {\tt P8R3~SOURCE} event type without sub-division into PSF types, acceptance weighted) at 1~GeV photon energy.
    }
    \label{fig:extension}
\end{figure}

\begin{figure}
    \includegraphics[width=\columnwidth]{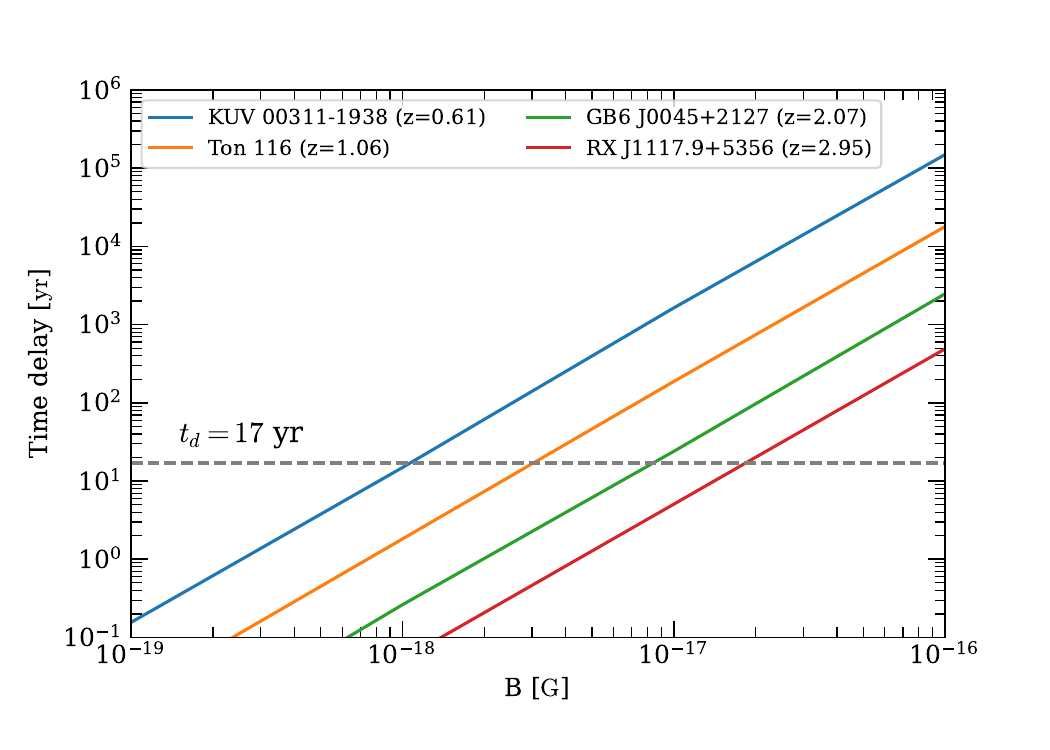}
    \caption{
        Median time delay of the ``echo'' photons in the $E_\gamma=0.25-0.35$~GeV energy range for the same sources as in Fig.\ref{fig:extension}. For a given IGMF strength the calculated delay varies from source to source due to its strong $t_d \sim (1+z)^{-5}$ redshift dependence~\citep{NeronovSemikoz09}. The horizontal dashed line marks the duration of the Fermi/LAT observations employed here.
    }
    \label{fig:tdelay}
\end{figure}

\CRBeam\ simulations enable direct estimates of the spatial extension and time delay of the ``echo'' signal in the presence of IGMF for the studied sources with their outputs recording the final position and direction of all the ``echo'' photons as they reach the observer. Based on them, the time delay was estimated here as the source-to-observer and the finial source-to-photon distances difference divided by the speed of light; the spatial extension was estimated as a 68\% quantile of the ``echo'' photons arrival direction offset with respect to the observer-to-source axis. The computed extension and time delay for selected sources, covering the redshift range of the AGNs in Tab.~\ref{tab:high-z-agns} and roughly representing the employed redshift bin edges, are shown in Figs.~\ref{fig:extension} and \ref{fig:tdelay} respectively, where they are compared with the Fermi/LAT point spread function and observational window duration. To account for the latter, expected ``echo'' extension was also recalculated using only photons arriving to the observer with the time delay less than 17~yr, as shown in Fig.~\ref{fig:extension} with dotted lines.

These calculations allow to estimate the sensitivity range to IGMF in terms of the ``echo'' time delay.
The relation between the intrinsic and ``echo'' emission energies, $E_\gamma \approx 0.8 (1+z)^2 (E_0 / \mathrm{1~TeV})^2~\mathrm{GeV}$, implies the bulk of the ``echo'' flux corresponding to the $0.1-1$~TeV emission probed by Fermi/LAT is realized at energies $E_\gamma \lesssim 1$~GeV. Comparing the corresponding time delay from the simulations with the duration of Fermi/LAT observations, shown in Fig.~\ref{fig:tdelay}, it follows the IGMF with the strength at least up to $B \sim 10^{-18} - 10^{-17}$~G would not be able to suppress the ``echo'' emission throughout the observations time window and can be probed with Fermi/LAT data provided the expected ``echo'' signal is strong enough.

\section{Variability of the selected sources}
\label{sect:variability}

\begin{figure*}
    \includegraphics[width=\linewidth]{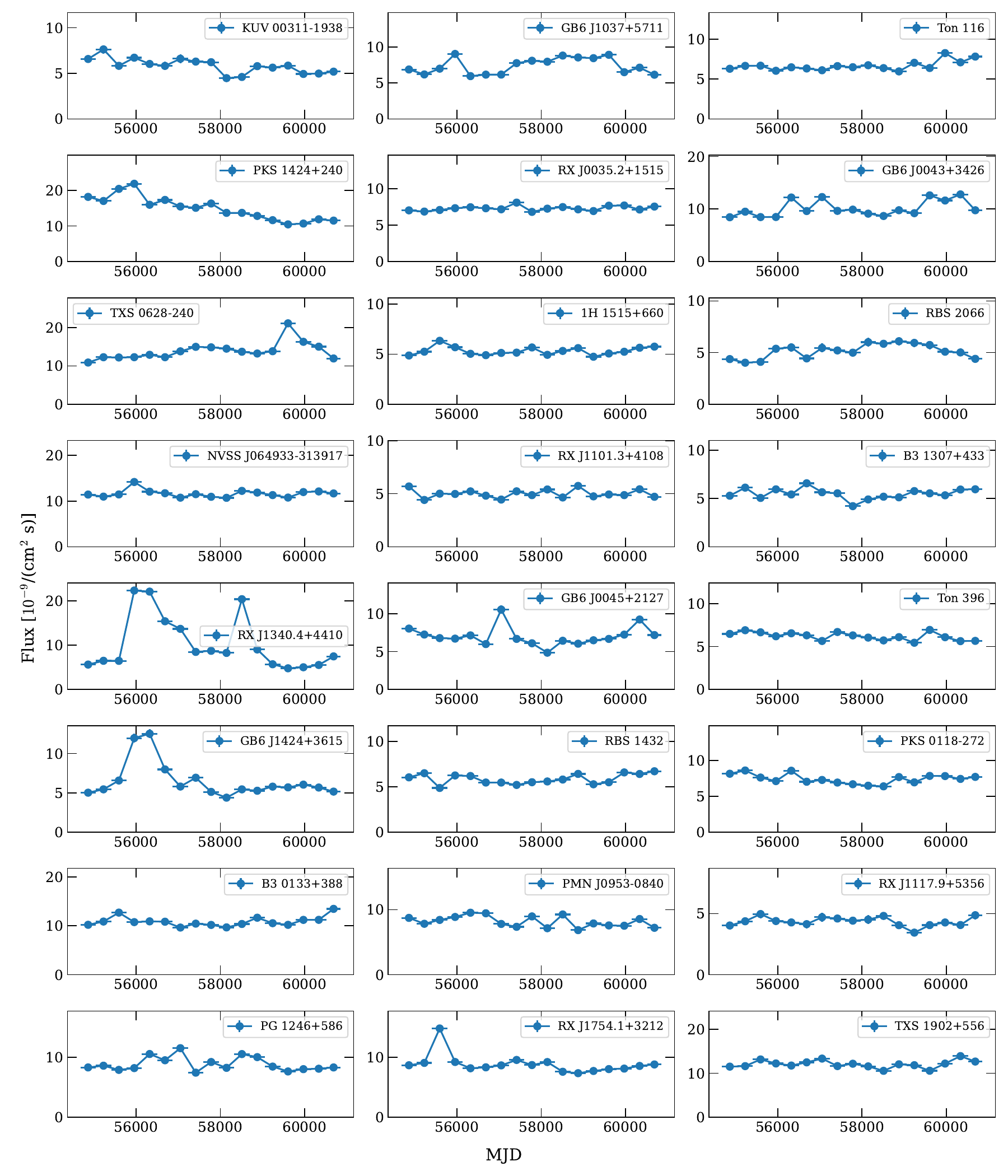}
    \caption{
        Light curves of the sources from Tab.~\ref{tab:high-z-agns} extracted above 1~GeV from Fermi/LAT data with yearly binning.
    }
    \label{fig:lightcurves}
\end{figure*}

The yearly-binned light curves of the sources in Tab.~\ref{tab:high-z-agns} were extracted using the aperture photometry approach from the $2^\circ$ circles around the catalogue source positions, selecting {\tt P8R3~SOURCE} events with energy above 1~GeV as described in the \Fermitools\ documentation\footnote{\url{https://fermi.gsfc.nasa.gov/ssc/data/analysis/scitools/aperture_photometry.html}}; these are shown in Fig.~\ref{fig:lightcurves}.

The impact of the source variability on the derived IGMF limits stems from the time-delayed nature of the ``echo''. Under the assumption of the constant (time-averaged) source flux, this may lead to a biased ``echo'' power estimate if a substantial fraction of the latter is accumulated at the beginning or an end of the Fermi/LAT observations time window. The relative deviation $F$ of the accumulated emission power from the constant in time assumption within an interval $[0;T]$ can be parametrized as

\begin{equation}
    \Delta F(t) = \frac{
        \int_0^t f(t) dt - \left< f \right> t
    }{
        \left< f \right> T
    }
\end{equation}

The maximum of $\Delta F(t)$ for each source estimates the amount of ``echo'' power with may have an additional time delay within $t_d \in [0;T]$ window with respect to that set by the assumed IGMF. For the sources studied here this value is $\Delta F_{max} \lesssim 5\%$ with exception of PKS~1424+240, GB6~J1424+3615 (both with $\Delta F_{max} \approx 9\%$) and RX~J1340.4+4410 ($\Delta F_{max} \approx 14\%$).

\bibliographystyle{elsarticle-num} 
\bibliography{references}

\end{document}